\documentclass[final,twocolumn]{elsarticle}

\usepackage{amssymb}
\usepackage{amsmath}
\usepackage{epsfig}
\usepackage{bm}
\usepackage{latexsym}
\usepackage{color}
\usepackage{epstopdf}

\journal{Physics Letters A}

\begin{document}

\begin{frontmatter}



\title{Controlling thermal conductivity  in harmonic chains with correlated mass and  bond disorder: Analytical approach}


\author{I.~F.~Herrera-Gonz\'alez}

\address{Decanato de Ingenier\'ias, UPAEP University, 21 Sur 1103, Barrio de Santiago, Puebla, Puebla, 72410, M\'exico}



\begin{abstract}
We investigate heat transport in one-dimensional harmonic chains with mass disorder and weak bond disorder, coupled at both ends to oscillator heat baths through weak impedance mismatches. The model incorporates correlations in mass disorder, in bond disorder, and between the two. We find that the scaling of thermal conductivity $\kappa$  with system size $N$ is determined solely by either mass disorder or bond disorder. This indicates that cross-correlations between the two types of disorder play no important role in the scaling behavior of $\kappa$. Consequently, by tuning the self-correlations, it is possible to control how the thermal conductivity scales with the system size. Such control could have potential applications in thermoelectric devices and thermal insulation technologies.  
\end{abstract}




\end{frontmatter}



\section{Introduction}
\label{sec1}
The primary goal of nonequilibrium statistical mechanics is to derive phenomenological laws from microscopic dynamics. One such law is Fourier’s law
, which states that the amount of heat transported per unit surface area and per unit time is proportional to the temperature gradient. The proportionality constant, known as thermal conductivity (denoted by $\kappa$), is an intrinsic property of the material and an intensive quantity. When energy conservation is taken into account, this law predicts a linear temperature profile under a small temperature bias along the direction of heat flow in the steady state. However, several numerical and theoretical studies on low-dimensional lattices have shown that $\kappa$ often depends on the system size and may even diverge in the thermodynamic limit, thereby violating Fourier{'}s law \cite{LLP03,D08,L15}.  Experimental evidence for this breakdown has been reported in boron nitride nanotubes \cite{COGMZ08}, nanowires \cite{YZL10}, in single-layer graphene \cite{XPWZ14}, and in other quasi-one dimensional systems \cite{LXXZL12}.

One of the earliest models investigated in the study of heat conduction is the homogeneous harmonic chain, where the thermal conductivity $\kappa$  scales with system size $N$ as $\kappa\sim N$ \cite{RLL1967}.  This anomalous scaling arises because phonons do not scatter in the isolated system. Thus, some sort of anharmonicity, disorder or other internal or external mechanisms are needed to introduce interaction among phonons and recover Fourier's law. Introducing uncorrelated mass disorder localizes all modes in the thermodynamic limit (Anderson localization), but low-frequency modes in finite chains still remain extended, leading  to an anomalous thermal conductivity \cite{MI70,RG71,CL1971,CL74,V79}. Moreover, the addition of anharmonicity \cite{lepri1997}  or external pinning potentials \cite{I73,RD08} alone is insufficient to restore normal conductivity.       

Finite thermal conductivity can still be achieved  through different mechanisms,
including the presence of periodic external potentials (as in the Frenkel–Kontorova model \cite{Frenkel1938,hu2005}), self-consistent inner reservoirs \cite{rich1975,bonetto2004}, energy-conserving stochastic noise \cite{dhar2011,landi2013,landi2014,palla2020}, and the combination of pinning potentials with anharmonic interparticle interactions \cite{dhar2008effect}. Other systems that obey Fourier’s law include the ding-a-ling \cite{casati1984}, ding-dong \cite{prosen1992,sano2001}, Lorentz gas \cite{alonso1999}, anharmonic chain with weak interparticle interactions \cite{Pereira2006}, and lattice billiard models \cite{gaspard2008}. A common characteristic of these systems is the absence of momentum conservation, which is generally regarded as a necessary condition for the validity of Fourier’s law. However, notable exceptions exist: momentum-conserving models such as coupled rotator chains \cite{giardina2000,gendelman2000}, graded-mass harmonic chains \cite{reich2013}, and two-dimensional lattices with missing bonds \cite{yang2002}, also display finite thermal conductivity. Similarly, momentum-conserving harmonic chains with strong bond disorder \cite{AOI18,AABOI19} or with mass disorder and suitable  thermal baths \cite{D01}  display the same behavior.  Nonetheless, rigorous results show that disordered harmonic chains with conserved momentum cannot sustain a linear temperature profile \cite{HK21}, and thus violate Fourier’s law.

The precise conditions under which Fourier’s law holds remain unresolved, but model studies have revealed the influence of momentum conservation, dimensionality, disorder, and nonlinearity on heat transport. In real materials such as alloys or nanotubes, however, disorder is rarely purely random and often exhibits correlations \cite{ikematsu2000,sebastian2024}. Such correlations play a crucial role in heat conduction and have recently gained attention for their potential in tuning thermal conductivity, with applications in thermoelectrics, insulation, and related technologies \cite{chaney2021,thebaud2023}. Since a general theoretical framework is still lacking, clarifying the one-dimensional case may provide insights into higher-dimensional systems. In addition, the harmonic approximation is particularly useful, as anharmonic effects become negligible at low temperatures

 Harmonic chains with specific disorder correlations can break Anderson localization and generate extended modes at discrete frequencies or within defined frequency intervals \cite{MCRL03,HIT10,JLM15,SRSSM19,ZZ19,HM23}. Disorder correlations  can also affect phonon transport \cite{ASML15,ZZ14,Y18}, influence clustering within the chain \cite{SZG17}, and enable direct control of thermal conductivity scaling in isotopically disordered chains \cite{ZLW15,HIT15,HM20}. However, existing studies have largely focused on either purely mass disorder, purely bond disorder, or their combination under specific correlation schemes. What remains lacking is a general theoretical framework describing the scaling of thermal conductivity when both types of disorder are simultaneously present with arbitrary correlations. Addressing this gap is particularly important for practical applications:in regular doping or in an isotopic doping, substituting an extrinsic atom at a lattice site alters not only the local mass but also the associated force constants. Specifically, elemental doping modifies the electronic structure and local chemical environment, thereby affecting the bond strengths of neighboring atoms, whereas isotopic doping primarily changes the nuclear mass, slightly shifting equilibrium bond lengths, though chemical bonding may also be affected \cite{fleming2014}. Consequently, such substitutions naturally induce correlations between mass and bond disorder \cite{FDCRMM21,NLKVC22}.

The purpose of this work is twofold. First, we provide a general framework for describing how the thermal conductivity scales with system size in harmonic chains containing correlated mass and weak bond disorder, applicable for arbitrary stationary correlations of the disorder. Second, we demonstrate how to construct tailored sequences of masses and spring constants that enable precise control of thermal conductivity scaling with system size.

\section{The model}
We consider a one-dimensional lattice of $N$ atoms, with nearest-neighbor harmonic interactions, described by the Hamiltonian  
\begin{equation}
H=\sum^N_{j=1} \frac{p^2_j}{2m_j}+\frac{1}{2} \sum^N_{i,j=1}K_{i,j}q_iq_j,
\label{dynamic}
\end{equation} 
where $q_n$ denotes the displacement of the mass $m_n$ with respect to its equilibrium position, and $p_n$ is its momentum. The force constant matrix $K_{i,j}$ is a tri-diagonal matrix defined as
\begin{equation}
K_{i,j}=(k_i+k_{i-1})\delta_{i,j}-k_i\delta_{i,j-1}-k_{i-1}\delta_{i,j+1}.
\end{equation} 
Here, $k_j$ denotes the coupling strength between the masses $m_{j}$ and $m_{j+1}$. In this model, $m_j$ and $k_j$ are correlated random variables with common mean values $m=\langle m_j\rangle$ and $k=\langle k_j\rangle$, and common variances $\sigma^2_m$ and $\sigma^2_k$, with  $\langle \cdots \rangle$ denoting disorder average. The boundary conditions are determined by the matrix elements $K_{1,j}$ and $K_{N,j}$, however, this will remain unspecified at this stage. 

Bond disorder is assumed to be weak
\begin{equation}
\frac{\sigma_k}{k}\ll1,
\end{equation} 
 which means small fluctuations of the elastic constants $k_i$ around their mean values. No such weak disorder condition is imposed on the random masses, as their effective fluctuations are small in the low-frequency limit. Therefore, to complete the description of the model, only the following binary correlators are needed  
\begin{eqnarray}
\label{corre}
\chi_1(l)&=&\frac{\langle \delta m_n \delta m_{n-l} \rangle}{\sigma^2_m}, \nonumber \\
\chi_2(l)&=&\frac{\langle \delta k_n \delta k_{n-l} \rangle}{\sigma^2_k},  \\
\chi_3(l)&=&\frac{\langle \delta m_n \delta k_{n-l} \rangle}{\langle \delta k_n \delta m_n\rangle}. \nonumber
\end{eqnarray} 
Where $\delta m_n=m_n-m$ ($\delta k_n=k_n-k$) denote the fluctuations of the random masses (bond strengths) around their mean value. The binary correlators depend solely on the relative index separation $l$ between any two elements of the sequences $\{m_n\}$ and $\{k_n\}$. This property implies that both sequences  are stationary in the weak sense. 
 
To analyse heat transport, we connect the ends of the disordered harmonic chain to two oscillator heat baths, each composed of harmonic oscillators with identical masses $m$ and identical bond strengths $k$. This configuration produces a small impedance mismatch between the disorder chain and the thermal baths. The left and right baths are at temperatures $T_L$ and $T_R$, respectively. Choosing $k_0=k_N=k$ in the matrix force $K_{i,j}$ imposes free boundary conditions which preserves the zero frequency mode. Alternative choices of $k_0$ and $k_N$ lead to pinned boundary sites; however, the present 
work considers only the free-boundary case.

The thermal conductance (defined as the ratio  of the steady state heat current in the two-terminal system to the temperature difference between the heat baths) is evaluated in the small temperature bias $T_L-T_R$ by \cite{AOI18,RK98} 
\begin{eqnarray}
G(N,\theta)=\frac{\hbar}{2\pi}\int^{\infty}_{0}\omega\frac{\partial f(\theta,\omega)}{\partial \theta} T(\omega) d\omega,
\label{flux}
\end{eqnarray}
where $T(\omega)$ is the transmission coefficient of phonons through the disordered chain,  $f(\theta,\omega)$ is the Bose-Einstein distribution
\begin{equation}
f(\theta,\omega)=\left(e^{\hbar\omega/\theta}-1\right)^{-1}
\end{equation} 
with $\theta=(T_L-T_R)/2$ the mean temperature between the two heat baths. The Boltzmann constant has set equal to 1.
\section{Localization length} 
The localization length $L_{\text{loc}}$ characterizes the spatial extend of an exponentially localized mode in an infinite system. Recently, a second-order perturbative expression for its inverse $\lambda=L^{-1}_{\text{loc}}$ has been obtained for harmonic chains with correlated mass and weak bond disorder. The expression reads as \cite{HM23}
\begin{equation}
\lambda=\frac{\tan^2\left(\frac{\mu}{2}\right)}{2}\left[\widetilde{\sigma}^2_mW_1(\mu)+\widetilde{\sigma}^2_kW_2(\mu)
+2\widetilde{\Delta}\cos\mu W_3(\mu)\right],
\label{final}
\end{equation}
where $\widetilde{\sigma}^2_m=\sigma^2_m/m^2$ and $\widetilde{\sigma}^2_k=\sigma^2_k/k^2$ are dimensionless variances for mass and bond disorder, respectively. $\widetilde{\Delta}=\left<\delta m_n \delta k_n\right>/km$ measures the cross-correlation at site $n$, and the parameter $\mu$ is related to the frequency $\omega$ through the expression 
\begin{eqnarray}
\omega^2=\frac{4k}{m}\sin^2\left(\frac{\mu}{2}\right),
\label{disper} 
\end{eqnarray}
which corresponds to the dispersion relation of the harmonic chain in the absence of disorder, with the lattice constant set to unity. The functions $W_i(\mu)$, $i=1,2,3$, are given by the Fourier transforms of the normalized binary correlators $\chi_i(l)$
\begin{eqnarray}
W_i(\mu)=1+2\sum^{\infty}_{l=1}\chi_i(l)\cos(2l\mu)
\label{spectral} 
\end{eqnarray}
Therefore, the first two of those functions ($i=1,2$) correspond to the power spectra of the fluctuation of random masses and random springs, respectively, whereas, the third one represents the interplay between the bond disorder and mass disorder.  $W_i(\mu)$ are pair functions of period $\pi$ that satisfy the following normalization conditions
 \begin{eqnarray}
\int^{\frac{\pi}{2}}_0 W_i(\mu)d\mu=\frac{\pi}{2} 
\label{normal}
\end{eqnarray}
 and the following inequality
\begin{eqnarray}
\widetilde{\sigma}^2_mW_1(\mu)+\widetilde{\sigma}^2_kW_2(\mu)
\ge 2|\widetilde{\Delta} W_3(\mu)\cos(\mu)|.
\label{ine}
\end{eqnarray} 
 The proof of this inequality is provided in the appendix. Notice that $\chi_i(l)$ are even functions of the index $l$, this can be proved directly if we invert Eq. (\ref{spectral}) to write $\chi_i(l)$ in terms of $W_i$.

In addition, expression (\ref{final}) is derived using a second-order perturbative approach, subject to  the following restrictions  
\begin{eqnarray}
4\left(\frac{\omega}{\omega_m}\right)^2\widetilde{\sigma}_m \ll 1 \ \ \text{and} \ \ \widetilde{\sigma}_k \ll 1,
\label{weak}
\end{eqnarray}
where $\omega_{m}=\sqrt{4k/m}$ is the upper limit of the frequency spectrum for the homogeneous chain. 

Since our analysis focuses on the low-frequency limit, the first restriction of Eq. (\ref{weak}) is automatically satisfied for practically any value of the standard deviation of the random masses. The second condition, however, establishes that bond disorder must be weak.

\section{Cross-correlations and thermal conductance}
When one considers the finite size chain attached to thermal reservoirs, only the low frequency modes will contribute to the heat conduction if the size of the chain is large enough. This can be seen clearly from Eq. (\ref{final}), where low frequency modes  have a larger localization length that goes to infinity at the zero frequency. Thus, there is a cut-off frequency $\omega_c$ defined by the equation
\begin{equation}
L_{\text{loc}}(\omega_c)=N;
\label{cut}
\end{equation}
Modes whose frequencies are above $\omega_c$ do no contribute to the heat flux because they  are localized if one compares $L_{\text{loc}}$ with the system size $N$. Thus, only the low frequency limit of Eq. (\ref{final}) must be considered, and one must identify the leading term in that limit from among the three terms of the r.h.s of that equation. The term that contains the information about cross-correlations cannot be the leading term because of the inequality (\ref{ine}).

For small impedance mismatch between the baths and the disordered chain,  the transmission coefficient for modes below $\omega_c$ will be roughly equal to one. In addition, if $N\gg1$, then $\omega_c<<1$. Thus, for long system sizes or high mean temperature $\theta$, one can compute the thermal conductance (\ref{flux}) in terms of the following series expansion \cite{AABOI19}

\begin{eqnarray}
\frac{G(N,\theta)}{g_0}=\frac{3}{\pi^2}\left(\frac{\omega_c}{\omega_\theta}\right)+O\left( \frac{\omega_c}{\omega_\theta} \right)^2,
\label{a2}
\end{eqnarray} 
where $\omega_T=\theta/\hbar$ is the thermal frequency that determines if a given frequency is populated with phonons or not (if $\omega<\omega_T$, then $\omega$ is populated), and $g_0=\pi^2\theta/3h$  is the quantum thermal conductance which is the universal value for  $G(N,\theta)$ in the limit $\theta \rightarrow 0$.

Equations (\ref{cut}) and (\ref{a2}) determine how the thermal conductance scales with  the system size. Thus, if we consider that the term containing cross-correlations cannot be the leading term in Eq. (\ref{final}) at low frequencies, then cross-correlations cannot determine how the thermal conductance scales with the system size. This statement is one of the main novel results of this work, since it is a general statement for any disordered harmonic chain under weakly stationary conditions, weak bond disorder and small impedance mismatch between heat baths and harmonic chain. 
\section{Scaling behavior of thermal conductance}
For one dimensional systems, the effective thermal conductivity $\kappa$ (here just called thermal conductivity) is defined as the ratio between the heat flux $J_N$ and the temperature gradient $\kappa=NJ_N/(T_R-T_L)$ \cite{LLP03}, and it is related to the thermal conductance through the equation $\kappa=NG$ 

In order to obtain the scaling behavior of $G$ (or $\kappa$), one must specify the low frequency dependence of the power spectra $W_1$ and $W_2$. The following choices correspond to the most common situation in correlated disordered systems \cite{NLKVC22}:
\begin{eqnarray}
W_i(\omega)\propto \omega^{\beta_i}, \quad \text{when} \quad \omega \rightarrow 0,
\label{low}
\end{eqnarray}
with $i=1,2$. The exponents $\beta_i$ must satisfy the following inequalities due to the normalization conditions (\ref{normal})
\begin{equation}
\beta_i>-1.
\label{ine2}
\end{equation} 
If one takes into account equations (\ref{cut}), (\ref{a2}) and (\ref{low}), we arrive at the following results for the thermal conductance and for the thermal conductivity:
\begin{eqnarray}
G\sim N^{\alpha}, \ \  \alpha=-\frac{1}{2+\beta}, 
\label{finalf}
\end{eqnarray} 

\begin{eqnarray}
\kappa\sim N^{\alpha'} \ \  \alpha'=\frac{1+\beta}{2+\beta},
\label{finalf1}
\end{eqnarray}
where $\beta=\text{min}(\beta_1,\beta_2)$ is the minimum value between $\beta_1$ and $\beta_2$. Here, only the minimum value of the exponents $\beta_i$ is taken into account due to the fact that the power spectrum associated with this value will be the leading term in Eq. (\ref{final}) for the low frequency regime. If no bond disorder exist, $\beta=\beta_1$, and the expression of $\kappa$ for correlated mass disorder is recovered \cite{HIT15}. Nonetheless, the expression obtained in that reference is derived using a classical heat current, whereas expressions (\ref{finalf}) and (\ref{finalf1}) are also valid  in the quantum regime. If only purely bond disorder is considered, $\beta=\beta_2$, in addition, if this disorder is uncorrelated, $\beta_2=0$, and $\kappa=1/2$, which coincides with previous results derived assuming uniform distribution or a power law distribution \cite{AABOI19}. However, the result (\ref{finalf}) is valid for any distribution under the weak bond disorder assumption (\ref{weak})

In addition, $\alpha>-1$ ($\alpha'>0$) which means that the thermal conductivity cannot decrease or be independent of the system size. These results come directly from the restrictions of the exponents $\beta_i$ (\ref{ine2}).

Eq. (\ref{finalf}) (or Eq. (\ref{finalf1})) is the second novel result of this work. In the next section, we will show  how disorder correlations can be manipulated to control the size scaling of the thermal conductance (or thermal conductivity) using those equations.

\section{Numerical results}
\label{Num}
In order to generate random sequences $\{m_n \}$ and $\{k_n\}$ with pre-defined power spectra $W_i$, $i=1,2$, we use the standard technique of creating colored noised from discrete white noise by using the following convolution operation \cite{HITT10}:
\begin{equation}
Z^{(i)}_n=\sum^{\infty}_{n'=-\infty} G_i(n')x^{(i)}_{n-n'}+\left<Z^{(i)}_n\right>,
\end{equation} 
with $x^{(i)}_n$ random variables with the properties $\left< x^{(i)}_n\right>=0$ and  $\left< x^{(1)}_nx^{(2)}_{n'}\right>=\delta_{n,n'}$. In addition, the following notation is introduced: $Z^{(1)}_n=m_n$ and $Z^{(2)}_n=k_n$. The modulation functions $G_i(n)$ are given in terms of the pre-defined power spectra $W_i$
\begin{equation}
G_i(n)=\frac{2}{\pi}\int^{\pi/2}_0\sqrt{\sigma_iW_i(\mu)}\cos(2\mu n)d\mu.
\end{equation}
Here, $\sigma_1=\sigma_m$, and $\sigma_2=\sigma_k$. To introduce cross-correlations between mass and bond disorder, the following transformation is used
\begin{eqnarray}
         \left(
                \begin{array}{r}
                x^{(1)}_{n}\\
                 x^{(2)}_{n}
                \end{array}
         \right)=  \begin{array}{rr}
  
        \left(
                \begin{array}{rr}
                 \cos \delta & \sin \delta\\
                 \sin \delta & \cos \delta 
                \end{array}
         \right)\left(\begin{array}{r}
                y^{(1)}_{n} \\
                y^{(2)}_{n}
                \end{array}\right),
  \end{array}
\label{map}
\end{eqnarray}
where $\left<y^{(i)}_n\right>=0$, $\left<y^{(i)}_ny^{(j)}_m\right>=\delta_{i,j}\delta_{m,n}$. In this way, $\left<x^{(1)}_nx^{(2)}_{n'}\right>=\sin(2\delta)\delta_{n,n'}$, and the parameter $\delta$ controls the degree of correlation between the random variables $m_n$ and $k_n$ : When $\delta=0$ no cross-correlations exist. $\delta=\pi/4$ corresponds to total correlations ($m_n/m=k_n/k$) and $ \delta=-\pi/4$ is associated with total anti-correlations ($\delta m_n/m=-\delta k_n/k$). Thus, the parameter $\delta$ lies in the interval $[-\pi/4,\pi/4]$.

With this particular choice of creating the random sequences, the term $W_3$ is given by
\begin{equation}
W_3(\mu)=\widetilde{\sigma}_m\widetilde{\sigma}_k\sqrt{W_1(\mu)W_2(\mu)}\sin(2\delta).
\end{equation}
In the numerical simulations, we work in units where $\hbar=1$ and $k=m=1$. We set $\widetilde{\sigma}^2_m=\widetilde{\sigma}^2_k=0.01$ and $\theta=0.1$. The thermal conductance is averaged over \textcolor{blue}{50} different realizations of the disorder. Error bars represent the standard deviation measuring the data dispersion around the mean. The transfer matrix method is used to calculate the transmission coefficient in Eq. (\ref{flux}). Additionally, to test numerically the results of the previous section, we work with the following power spectra   
\begin{eqnarray}
W_i(\mu)=(\beta_i+1)\left(\frac{2|\mu|}{\pi}\right)^{\beta_i}, \ \ i=1,2.
\end{eqnarray}
Therefore, $W_3\propto \mu^{(\beta_1+\beta_2)/2}$, and this term cannot be the leading term in Eq. (\ref{final}) for low frequencies, which it is consistent with our previous analysis. 

\begin{figure}[t]
\includegraphics[width=\columnwidth,scale=0.9]{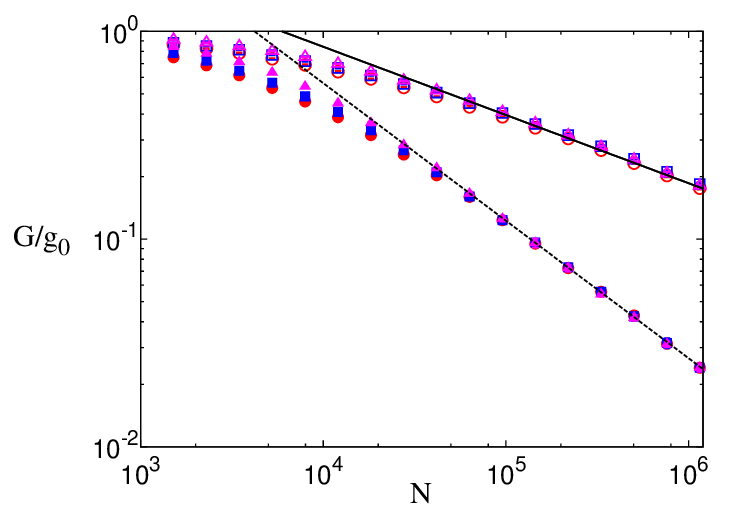}
\caption{Dimensionless thermal conductance as a function of the system size. Different point styles represent numerical data for different values of $\delta$: circles ($\delta=\pi/4$), squares ($\delta=0$), triangles ($\delta=-\pi/4$). Filled points (empty points) correspond to numerical data with $\beta_1=3$ ($\beta_1=1$) and $\beta_2=-0.5$ ($\beta_2=4$) . Continuous line (dashed line) depicts  the best fit of the numerical data to the function $f(N) = aN^\alpha$ with $\alpha=-0.66\pm 0.01$ ($\alpha=-0.33\pm 0.01$). 
The size of the error bars are less than the point size}
\label{Fig1}
\end{figure}

Figure \ref{Fig1} shows the behavior of the thermal conductance (rescaled to the quantum thermal conductance $g_0$)  to the system size  for three different values of $\delta$ and two different values of the exponents $\beta_1$ and $\beta_2$, we see that the thermal conductance is independent of cross-correlations when the system size is large enough ($N\approx10^ 5$). For $\beta_1=1$ and $\beta_2=4$, the scaling behavior of $G$ is determined by purely mass disorder, and  Eq. (\ref{finalf}) predicts an exponent $\alpha=-1/3$ which agrees with the fitting of the numerical data (see the continuous line of the figure). When  $\beta_1=3$ and $\beta_2=-0.5$ only bond disorder determines the scaling behavior of $G$ with an exponent $\alpha=-2/3$, this is also in complete agreement with the numerical simulation (see the dashed line of the figure)

\section{Conclusion}
\label{con}
We investigate heat conduction in harmonic chains with correlated mass and weak bond disorder coupled to oscillator heat baths. The scaling of thermal conductivity is determined solely by either mass or bond disorder, making cross-correlations irrelevant. For disorder with low-frequency power-law spectra, we derive the system-size dependence of conductivity. Our results, valid in both quantum and classical regimes, indicate that doping may fail to control heat transport, as doping-induced bond disorder can control thermal  conductivity scaling, even when its strength is weak. Finally, we present a method to design mass and bond sequences that realize targeted thermal conductivity scaling, with potential applications in insulators and semiconductors.




\appendix
\section{}
\label{appex}
We define the following two quantities $ D_{\pm}$ as
\begin{equation}
D_{\pm}=\lim_{N\rightarrow \infty}\frac{1}{2N}\left<\left|\sum^{N}_{n=-N}(\widetilde
{\delta m_n}\pm\widetilde{\delta k_n})e^{2in\mu} \right|^2\right>,
\label{Ap1}
\end{equation}
where $\widetilde{\delta m_n}=\delta m_n/m$ and $\widetilde{\delta k_n}=\delta k_n/k$ are the normalized fluctuations of the random masses and random springs, respectively. If one uses the fact that $\widetilde{\delta m_n}$, $\widetilde{\delta k_n}$ are stationary successions in the weak sense, and the normalized binary correlators $\chi_i(l)$ are pair functions of the index $l$, after some algebraic manipulations, one obtains 

\begin{eqnarray*}
D_{\pm}=\widetilde{\sigma}^2_m W_1(\mu)\pm 2\widetilde{\Delta}W_3(\mu)+\widetilde{\sigma}^2_k W_2(\mu).
\end{eqnarray*}

By definition (\ref{Ap1}), $D_{\pm}\ge 0$, thus  the following two inequalities are obtained

\begin{equation}
\widetilde{\sigma}^2_m W_1(\mu)+\widetilde{\sigma}^2_k W_2(\mu)\ge \pm 2\widetilde{\Delta}W_3(\mu).
\end{equation}

Both inequalities imply that

\begin{eqnarray}
\widetilde{\sigma}^2_mW_1(\mu)+\widetilde{\sigma}^2_kW_2(\mu)
\ge 2|\widetilde{\Delta} W_3(\mu) \cos(\mu)|.
\end{eqnarray} 

\section*{Acknowledgments}
I. F. Herrera-Gonz\'alez thanks UPAEP University for financial support






\bibliographystyle{elsarticle-num}
\bibliography{references1}

@article{LLP03,
	author={Lepri, Stefano and Livi, Roberto and Politi, Antonio},
	journal={Phys. Rep.},
	volume={377},
	pages={1},
	year={2003},
	publisher={Elsevier}
}

@article{D08,
	author={Dhar, Abhishek},
	journal={Adv. Phys.},
	volume={57},
	pages={457},
	year={2008},
	publisher={Taylor \& Francis}
}

@book{L15,
	title={Thermal transport in low dimensions},
	author={Lepri, Stefano and Livi, R and Politi, A},
	volume={921},
	year={2016},
	publisher={Springer, Berlin}
}

@article{COGMZ08,
	author={Chang, Chih-Wei and Okawa, David and Garcia, Henry and Majumdar, Arunava and Zettl, Alex},
	journal={Phys. Rev. Lett.},
	volume={101},
	pages={075903},
	year={2008},
	publisher={APS}
}

@article{YZL10,
	author={Yang, Nuo and Zhang, Gang and Li, Baowen},
	journal={Nano Today},
	volume={5},
	pages={85},
	year={2010},
	publisher={Elsevier}
}

@article{XPWZ14,
	author={Xu, Xiangfan and Pereira, L F C and Wang, Yu and Wu, Jing and Zhang, Kaiwen and Zhao, Xiangming and Bae, Sukang and  Bui, C T and Xie, Rongguo and Thong, J T L and Honk, B H and Loh, K P and Donadio, D and Li, B and \"Ozyilmaz, B},
	journal={Nat. Commun},
	pages={3689},
          volume={5},
	year={2014},
	publisher={Nature Publishing Group UK London}
}

@article{LXXZL12,
	author={Liu, Sha and Xu, X F and Xie, R G and Zhang, Gang and Li, B W},
	journal={Eur. Phys. J. B},
	volume={85},
	pages={337},
	year={2012},
	publisher={Springer}
}

@article{RLL1967,
	author={Rieder, Z and Lebowitz, J L and Lieb, E},
	journal={J. Math. Phys},
	volume={8},
	pages={1073},
	year={1967},
	publisher={American Institute of Physics}
}

@article{MI70,
	author={Matsuda, Hirotsugu and Ishii, Kazushige},
	journal={Prog. Theor. Phys. Suppl.},
	volume={45},
	pages={56},
	year={1970},
	publisher={Oxford Academic}
}

@article{RG71,
	author={Rubin, Robert J and Greer, William L},
	journal={J. Math. Phys.},
	volume={12},
	pages={1686},
	year={1971},
	publisher={American Institute of Physics}
}

@article{CL1971,
	author={Casher, A and Lebowitz, J L},
	journal={J. Math. Phys.},
	volume={12},
	pages={1701},
	year={1971},
	publisher={American Institute of Physics}
}

@article{CL74,
	author={O'Connor, A J and Lebowitz, J L},
	journal={J. Math. Phys.},
	volume={15},
	pages={692},
	year={1974},
	publisher={American Institute of Physics}
}

@article{V79,
	author={Verheggen, Theo},
	journal={J. Math. Phys.},
	volume={68},
	pages={69},
	year={1979},
	publisher={Springer}
}

@article{lepri1997,
  author={Lepri, Stefano and Livi, Roberto and Politi, Antonio},
  journal={Phys. Rev. Lett.},
  volume={78},
  pages={1896},
  year={1997},
  publisher={APS}
}

@article{I73,
	author={Ishii, Kazushige},
	journal={Prog. Theor. Phys. Suppl.},
	volume={53},
	pages={77},
	year={1973},
	publisher={Oxford Academic}
}

@article{RD08,
	author={Roy, Dibyendu and Dhar, Abhishek},
	journal={Phys. Rev. E},
	volume={78},
	pages={051112},
	year={2008},
	publisher={APS}
}

@article{Frenkel1938,
  author  = {Frenkel, Y. and Kontorova, T.},
  journal = {Zh. Eksp. Teor. Fiz.},
  volume  = {8},
  year    = {1938},
  pages   = {1340},
  language = {Russian}
}

@article{hu2005,
  author={Hu, Bambi and Yang, Lei},
  journal={Chaos},
  volume={15},
  year={2005},
  pages   = {015119},
  publisher={AIP Publishing}
}

@article{rich1975,
  author={Rich, M and Visscher, William M},
  journal={Phys. Rev. B},
  volume={11},
  pages={2164},
  year={1975},
  publisher={APS}
}

@article{bonetto2004,
  author={Bonetto, Federico and Lebowitz, Joel L and Lukkarinen, Jani},
  journal={J. Stat. Phys.},
  volume={116},
  pages={783},
  year={2004},
  publisher={Springer}
}

@article{dhar2011,
  author={Dhar, Abhishek and Venkateshan, K and Lebowitz, J L},
  journal={Phys. Rev. E},
  volume={83},
  pages={021108},
  year={2011},
  publisher={APS}
}

@article{landi2013,
  author={Landi, Gabriel T and de Oliveira, M{\'a}rio J},
  journal={Phys. Rev. E},
  volume={87},
  pages={052126},
  year={2013},
  publisher={APS}
}

@article{landi2014,
  author={Landi, Gabriel T and de Oliveira, M{\'a}rio J},
  journal={Phys. Rev. E},
  volume={89},
  pages={022105},
  year={2014},
  publisher={APS}
}

@article{palla2020,
  author={Palla, Pier Luca and Patera, Giuseppe and Cleri, Fabrizio and Giordano, Stefano},
  journal={Phys. Scr.},
  volume={95},
  pages={075703},
  year={2020},
  publisher={IOP Publishing}
}

@article{dhar2008effect,
  author={Dhar, Abhishek and Lebowitz, J L},
  journal={Phys. Rev. Lett.},
  volume={100},
  pages={134301},
  year={2008},
  publisher={APS}
}

@article{casati1984,
  author={Casati, Giulio and Ford, Joseph and Vivaldi, Franco and Visscher, William M},
  journal={Phys. Rev. Lett.},
  volume={52},
  pages={1861},
  year={1984},
  publisher={APS}
}

@article{prosen1992,
  author={Prosen, Toma{\v{z}} and Robnik, Marko},
  journal={J. Phys. A: Math. Gen.},
  volume={25},
  pages={3449},
  year={1992},
  publisher={IOP Publishing}
}

@article{sano2001,
  author={Sano, Mitsusada M and Kitahara, Kazuo},
  journal={Phys. Rev. E},
  volume={64},
  pages={056111},
  year={2001},
  publisher={APS}
}

@article{alonso1999,
  author={Alonso, Daniel and Artuso, Roberto and Casati, Giulio and Guarneri, Italo},
  journal={Phys. Rev. Lett.},
  volume={82},
  pages={1859},
  year={1999},
  publisher={APS}
}

@article{Pereira2006,
  author = {Pereira, Emmanuel and Falcao, Ricardo},
  journal = {Phys. Rev. Lett.},
  volume = {96},
  pages = {100601},
  year = {2006},
  publisher = {American Physical Society},
}

@article{gaspard2008,
  author={Gaspard, Pierre and Gilbert, Thomas},
  journal={New J. Phys.},
  volume={10},
  pages={103004},
  year={2008},
  publisher={IOP Publishing}
}

@article{giardina2000,
  author={Giardina, Cristian and Livi, Roberto and Politi, A and Vassalli, M},
  journal={Phys. Rev. Lett.},
  volume={84},
  pages={2144},
  year={2000},
  publisher={APS}
}

@article{gendelman2000,
  author={Gendelman, O V and Savin, A V},
  journal={Phys. Rev. Lett.},
  volume={84},
  pages={2381},
  year={2000},
  publisher={APS}
}

@article{reich2013,
  author={Reich, K V},
  journal={Phys. Rev. E},
  volume={87},
  pages={052109},
  year={2013},
  publisher={APS}
}

@article{yang2002,
  author={Yang, Lei},
  journal={Phys. Rev. Lett.},
  volume={88},
  pages={094301},
  year={2002},
  publisher={APS}
}

@article{AOI18,
	author={Amir, Ariel and Oreg, Yuval and Imry, Yoseph},
	journal={Europhys. Lett.},
	volume={124},
	pages={16001},
	year={2018},
	publisher={IOP Publishing}
}

@article{AABOI19,
	author={Ash, Biswarup and Amir, Ariel and Bar-Sinai, Yohai and Oreg, Yuval and Imry, Yoseph},
	journal={Phys. Rev. B},
	volume={101},
	pages={121403},
	year={2020},
	publisher={APS}
}

@article{D01,
	author={Dhar, Abhishek},
	journal={Phys. Rev. Lett.},
	volume={86},
	pages={5882},
	year={2001},
	publisher={APS}
}

@article{HK21,
	author={Hattori, Kiminori and Kumatoriya, Shohei},
	journal={J. Phys. Soc. Jpn.},
	volume={90},
	pages={114009},
	year={2021},
	publisher={The Physical Society of Japan}
}

@article{ikematsu2000,
  author={Ikematsu, Yoichi and Shindo, Daisuke and Oikawa, Tetsuo},
  journal={Mater. Trans. JIM.},
  volume={41},
  pages={238},
  year={2000},
  publisher={The Japan Institute of Metals}
}

@article{sebastian2024,
  author={Sebastian, Finn L and Settele, Simon and Li, Han and Flavel, Benjamin S and Zaumseil, Jana},
  journal={Nanoscale Horiz.},
  volume={9},
  pages={2286},
  year={2024},
  publisher={Royal Society of Chemistry}
}

@article{chaney2021,
  author={Chaney, D. and Castellano, A. and Bosak, A. and Bouchet, J. and Bottin, F. and Dorado, B. and Paolasini, L. and Rennie, S. and Bell, C. and Springell, R. and Lander, G. H},
  journal={Phys. Rev. Mater.},
  volume={5},
  pages={035004},
  year={2021},
  publisher={APS}
}

@article{thebaud2023,
  author={Th{\'e}baud, Simon and Lindsay, Lucas and Berlijn, Tom},
  journal={Phys. Rev. Lett.},
  volume={131},
  pages={026301},
  year={2023},
  publisher={APS}
}

@article{MCRL03,
	author={de Moura, F. A. B. F. and Coutinho-Filho, M. D. and Raposo, E. P. and Lyra, M. L.},
	journal={Phys. Rev. B},
	volume={68},
	pages={012202},
	year={2003},
	publisher={APS}
}

@article{HIT10,
	author={Herrera-Gonz{\'a}lez, I F and Izrailev, F M and Tessieri, L},
	journal={Europhys. Lett. },
	volume={90},
	pages={14001},
	year={2010},
	publisher={IOP Publishing}
}

@article{JLM15,
	author={J{\'u}nior, M P S and Lyra, M L and de Moura, F A B F},
	journal={Acta Phys. Pol. B},
	volume={46},
	pages={1247},
	year={2015}
}

@article{SRSSM19,
	author={ da Silva, L D   and Ranciaro-Neto, A and Sales, Messias O and dos Santos, J L L  and de Moura, F A B F},
	journal={An. Acad. Bras. Cienc.},
	volume={91},
	pages={e20180114},
	year={2019},
	publisher={SciELO Brasil}
}

@article{ZZ19,
	author={Zhai, Jianxiong and Zhang, Qingyun and Cheng, Zihan and Ren, Jie and Ke, Youqi and Li, Baowen},
	journal={Phys. Rev. B},
	volume={99},
	pages={195429},
	year={2019},
	publisher={APS}
}

@article{HM23,
	author={Herrera-Gonz{\'a}lez, I F and M{\'e}ndez-Berm{\'u}dez, J A},
	journal={Phys. Rev. E},
	volume={107},
	pages={034108},
	year={2023},
	publisher={APS}
}

@article{ASML15,
	author={de Albuquerque, S S and dos Santos, J L L and de Moura, F A B F and Lyra, M L},
	journal={J. Phys. Condens. Matter },
	volume={27},
	pages={175401},
	year={2015},
	publisher={IOP Publishing}
}

@article{ZZ14,
	author={Ong, Zhun-Yong and Zhang, Gang},
	journal={Phys. Rev. B},
	volume={90},
	pages={155459},
	year={2014},
	publisher={APS}
}

@article{Y18,
	author={H. S. Yamada},
	journal={Chaos, Solitons \& Fractals},
	volume={113},
	pages={178},
	year={2018},
	publisher={Elsevier}
}

@article{SZG17,
	author={Savin, Alexander V and Zolotarevskiy, Vadim and Gendelman, Oleg V},
	journal={Phys. Lett. A},
	volume={381},
	pages={145},
	year={2017},
	publisher={Elsevier}
}

@article{ZLW15,
	author={Zakeri, Sepideh S and Lepri, Stefano and Wiersma, Diederik S},
	journal={Phys. Rev. E},
	volume={91},
	pages={032112},
	year={2015},
	publisher={APS}
}

@article{HIT15,
	author={Herrera-Gonz{\'a}lez, I F and Izrailev, F M and Tessieri, L},
	journal={Europhys. Lett.},
	volume={110},
	pages={64001},
	year={2015},
	publisher={IOP Publishing}
}

@article{HM20,

	author={Herrera-Gonz{\'a}lez, I F and M{\'e}ndez-Berm{\'u}dez, J A},
	journal={Phys Lett. A},
	volume={384},
	pages={126380},
	year={2020},
	publisher={Elsevier}
}

@article{fleming2014,
  author={Fleming, Donald G and Manz, J{\"o}rn and Sato, Kazuma and Takayanagi, Toshiyuki},
  journal={Angew. Chem. Int. Ed.},
  volume={53},
  pages={13706},
  year={2014},
  publisher={Wiley Online Library}
}

@article{FDCRMM21,
	author={Fava, Mauro and Dongre, Bonny and Carrete, Jes\'us and van Roekeghem, Ambroise and Madsen, Georg K. H. and Mingo, Natalio},
	journal={Phys. Rev. B},
	volume={103},
	pages={174112},
	year={2021},
	publisher={APS}
}

@article{NLKVC22,
	author={Neverov, Vyacheslav D and Lukyanov, Alexander E and Krasavin, Andrey V and Vagov, Alexei and Croitoru, Mihail D},
	journal={Commun. Phys.},
	volume={5},
	pages={177},
	year={2022},
	publisher={Nature Publishing Group UK London}
	}

@article{RK98,
	author={Rego, Luis G. C. and Kirczenow, George},
	journal={Phys. Rev. Lett.},
	volume={81},
	pages={232},
	year={1998},
	publisher={APS}
	}

@article{HITT10,
	author={Hernandez Herrej{\'o}n, J C and Izrailev, F M and  Tessieri, L},
	journal={Physica E},
	volume={42},
	pages={2203},
	year={2010}
	}

	
\end{document}